# Probing Elastic Isotropy in Entropy Stabilized Transition Metal Oxides: Experimental Estimation of Single Crystal Elastic Constants from Polycrystalline Materials

Lalith Kumar Bhaskar[1,2,4*], Niraja Moharana[1,3], Hendrik Holz[4], Rajaprakash Ramachandramoorthy[4], K.C. Hari Kumar[2,3], Ravi Kumar[1,2]*

[1]Laboratory for High Performance Ceramics, Department of Metallurgical and Materials Engineering, Indian Institute of Technology Madras (IIT Madras), Chennai, 600036, India.

[2]Research Center on Ceramic Technologies for Futuristic Mobility, Indian Institute of Technology, Madras (IIT Madras), Chennai 600036, India.

[3]CALPHAD Lab, Department of Metallurgical and Materials Engineering, Indian Institute of Technology Madras (IIT Madras), Chennai 600036, India

[4]Max-Planck-Institut for Sustainable Materials, Max-Planck-Straße 1, 40237 Düsseldorf, Germany

*Corresponding authors – l.bhaskar@mpie.de; nvrk@iitm.ac.in

Abstract:

Single Crystal Elastic Constants (SECs) are pivotal for understanding material deformation, validating interatomic potentials, and enabling crucial material simulations. The entropy stabilized oxide showcases intriguing properties, underscoring the necessity for the determination of precise SECs to establish reliable interatomic potential and unlock its full potential using simulations. This study presents an innovative methodology for estimating SECs from polycrystalline materials, requiring only two diffraction elastic constants and isotropic elastic constants for crystals with cubic symmetry. Validation using phase-pure nickel demonstrated good agreement with existing literature values, with a maximum 11.5% deviation for $C_{12}$ values. Extending the methodology to [(MgNiCoCuZn)O], SECs were calculated: 219 GPa for $C_{11}$, 116 GPa for $C_{12}$, and 51 GPa for $C_{44}$. Comparison with literature-reported values from DFT calculations revealed a significant divergence, ranging from 25% to 59% in the bulk and shear modulus calculated using the Voigt-Reuss-Hill average. To comprehend this disparity, we conducted DFT calculations and thoroughly examined the factors influencing these values. This study not only introduces a straightforward and dependable SEC estimation methodology but also provides precise experimental SEC values for [(MgNiCoCuZn)O] entropy stabilized oxides at ambient conditions, crucial for developing accurate interatomic potentials in future research.

1. Introduction:

Advancements in materials science have paved the way for the development of novel materials with extraordinary properties and applications. Over time, numerous novel materials have emerged, yet a distinct group known as "high entropy alloys" has garnered significant interest from both researchers and industries alike. These materials possess a distinctive compositional space that enables the incorporation of four or more elements in nearly equal atomic ratios, effectively maximizing the configurational entropy. The ability to tailor these alloys for specific targeted applications has ignited a surge of research in this field [1,2].

In a pioneering effort to extend the concept of high entropy alloys to ceramic systems, researchers synthesized a novel multi-component oxide containing transitional metal ions Mg, Ni, Cu, Zn, and Co, resulting in a single-phase rock salt structure called "Entropy Stabilized Oxides" (ESO) [3]. This led to a surge of research with ceramics systems of other compositions and crystal structures as well [4–7]. Though numerous high entropy ceramic systems were synthesized, the transition metal ion stabilized rock salt structure [(MgNiCoCuZn)O] is of particular interest because of the high potential exhibited by this material system for various properties ranging from high Li-ion conductivity to low thermal conductivity [8–12]. Some of the properties were attributed to the Jahn-Teller distortions introduced into the lattice due to the presence of $Cu^{2+}$ ions [13,14]. Among the exhibited properties, the most intriguing and significant one is the elastic isotropy displayed by this material [15]. Pitike $et\ al.$ utilized first principles simulations to demonstrate that the non-magnetic form of [(MgNiCoCuZn)O] exhibits a remarkably high degree of isotropy, with a Zener ratio ($A^Z$) equal to 0.98 and for the paramagnetic and antiferromagnetic phase of this composition a value of 1.12 and 1.42 respectively was observed [15]. Very recently, using a diamond anvil cell *in situ* in a synchrotron beamline the mechanical behaviour of [(MgNiCoCuZn)O] powders was investigated under extreme pressure upto 40.1 GPa using radial X-ray diffraction [16]. In that work, the evolution of SECs was studied as a function of applied pressure. The calculation of SECs commenced at a minimum pressure of 1.2 GPa, and there are no available experimental studies that have determined SECs under ambient conditions. In these experiments, the Zener ratio was also computed, yielding a value of approximately 4.9 at 1.2 GPa, which then rapidly reduces to around 1 for pressures exceeding 21.4 GPa. The Zener ratio obtained through experimentation stands in stark contrast to the values derived from first principles calculations. Though the experimental measurements were conducted at room temperature and starting from 1.2 GPa pressure, while the first principle calculations were performed at 0 K and atmospheric pressures. To resolve this uncertainty, an extensive experimental study becomes essential to estimate the single crystal elastic constants (SECs) of [(MgNiCoCuZn)O] at ambient conditions and thereby determine the Zener ratio. With its remarkably low thermal conductivity of ~ 3 $Wm^{-1}K^{-1}$ [9,12], comprehending the elastic anisotropy of this material holds the potential for far-reaching applications.

The most common techniques used to measure SECs are resonant acoustic spectroscopy (RUS) and the Brillouin scattering and in both these techniques, sufficiently large single crystals are required [17]. But for this high entropy oxide, it is quite challenging to grow a single crystal of sufficiently large length and of the same composition as a polycrystalline counterpart. An alternative method for estimating SECs involves in situ loading of polycrystalline samples in a synchrotron or laboratory X-ray diffractometer, which allows the calculation of diffraction elastic constants (DECs). Using the DECs in conjunction with a micromechanical model and by employing least square regression, SECs could be obtained [18–23,23–25].

In this study, we present a novel methodology that combines uniaxial loading of samples in a custom-built loading fixture *in situ* in a laboratory X-ray diffractometer along with ultrasonic resonant frequency testing to estimate SECs from polycrystalline materials. Initially, the methodology was tested and validated on pure polycrystalline nickel to extract SECs successfully. Subsequently, the approach was extended to the [(MgNiCoCuZn)O] system. The choice of conducting initial experiments on pure nickel is based on its identical crystal structure with the [(MgNiCoCuZn)O] system. Moreover, SECs for nickel are extensively documented in the literature [26,27], ensuring a reliable validation of the proposed methodology. As a secondary validation, orientation-dependent elastic constants are computed using the experimentally obtained SECs of [(MgNiCoCuZn)O]. These computed elastic constants are then compared with the elastic constants experimentally obtained from nanoindentation mapping conducted on crystals with different orientations. Finally, the experimental SECs results are used to revisit the first principles calculations, leading to a comprehensive understanding of the correlation between phase composition and elastic anisotropy.

## 2. Materials and Methods:

### 2.1 Synthesis and Characterization

ESO of composition [(MgNiCoCuZn)O] was synthesized using solution combustion synthesis (SCS) [28]. The synthesized SCS powders are consolidated into pellets using spark plasma sintering (SPS) (Sumitomo Coal Mining Co. Ltd, Japan) in vacuum. The sintering of SCS powder particles was carried out in a graphite die of diameter 20 mm. The sintering process was carried out at 950 °C with a holding time of 5 min and at a heating rate of 100 °C/min. The applied pressure was increased at a rate of 10 MPa/min and reached a maximum value of 50 MPa at 500 °C. Thereafter, the pressure was kept constant till the end of the sintering time.

The compression samples were of dimensions $3\ mm\ \times 3\ mm\ \times 5\ mm.$ The thickness of the pellets were initially reduced from 5 mm to 3 mm using a low speed diamond cutting machine from Struers Ltd., USA. A low concentration diamond wafering blade from Chennai Metco Pvt Ltd., India, operating at 500 rpm was used for slicing. After reducing the thickness, the uniaxial compression samples were extracted from the pellet.

The microstructures, morphological features and elemental composition/mapping of the samples were determined using SEM from ThermoFischer - Apreo S (USA) coupled with energy-dispersive X-ray spectroscopy (EDS) analysis from EDAX Inc (USA). For texture analysis, the materials were subjected to standard metallographic polishing conditions. Samples were polished progressively with finer grade SiC emery paper followed by diamond polishing (0.25 μm) to get a mirror-finished surface without any scratches. As a final step the sample was ion milled for 15 mins at 5 kV using a precision *etching* coating system (*PECS) from* Gatan Inc.,(USA). EBSD images were acquired using Apreo S, Thermofischer (USA) equipped with an electron-back scattered attachment (high speed velocity™ EBSD camera, 3000 FPS), operating at an accelerating voltage of 30 kV. A scanning area of 600 μm × 450 μm and a step size of 1 μm was used. The microscopic images were analyzed using MTEX 5.6.0 in MATLAB R2018b software and the standard clean-up procedures such as confidence-index standardization and neighbor-orientation correlation were used. X-ray diffractograms were obtained using Bruker D8 Discover (Germany), operating at a voltage of 35 kV and a current of 30 mA. The diffractograms were obtained using Co $K_\alpha$ radiation in the 2θ range of 30° - 100° with a step size of 0.01˚ and scan speed per step of 1 s.

## 2.2 Estimation of mechanical properties

Nanoindentation mapping was carried out using G200, KLA Corporation (USA), with a standard Berkovich using the continuous stiffness method (CSM). The maximum penetration depth was set to 1000 nm at a constant strain rate of 0.01 $s^{-1}$. The CSM frequency was set to 45 Hz with an oscillation amplitude of 2 nm. A grid of 10 by 10 indents was conducted, totalling 100 indents. The spacing between each indent was 30μm. Post-nanoindentation EBSD of the sample was carried out to determine the orientation-dependent reduced elastic modulus. The reduced elastic modulus ($E_r$) was determined using the Oliver–Pharr method [29].

The Young's modulus, shear modulus and Poisson's ratio were obtained using ultrasonic resonant frequency testing method using a transducer and spectrum analyzer from KEYSIGHT Technologies (USA). The measurement was performed using a probe frequency of 2.25/5 MHz, energy of 50 to 100 μJ, damping of 25 Ohm and a gain of 15 to 40 dB. The standard equations were used to calculate the isotropic elastic constants (Young's modulus (E), shear modulus (G), bulk modulus (K) and Poisson's ratio ($v$)) from the measured longitudinal and shear velocities.

### 2.3 Estimation of diffraction elastic constants (DECs)

For the estimation of SECs both DECs and elastic constants obtained from ultrasonic resonant frequency testing are required. In brief, DECs are acquired by measuring lattice strain in the sample through the principles of X-ray diffraction during in situ loading. The lattice strain is

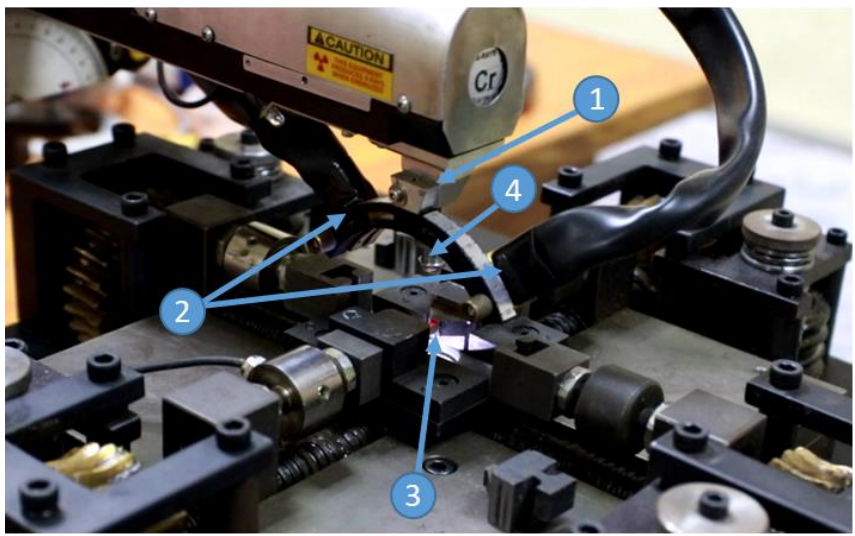

1- X-ray source

2- Detectors

3- Uniaxial compression coupon

4- Collimator

Figure 1        shows the close-up view of the setup for *in situ* uniaxial compression loading of the samples measured using *modified Chi mode*, more details about the method can be found in the supplementary information in section S1.

The measurement conditions to extract DECs are outlined as follows. The experimental setup for measuring DECs is shown in Figure 1. The laboratory X-ray diffractometer is from Proto iXRD (Canada) and the machine used for compression loading is a custom-built loading fixture and additional details about it can be found elsewhere [30]. To measure DECs, uniaxial compression coupons of dimension $3 \times 3 \times 5 \; mm^3$ were subjected to *in situ* loading. All the uniaxial loading experiments were within the elastic region of the samples. A collimator of 2 mm diameter was used to ensure that the X-rays fell on the gauge section of the sample. In total, 10 "β – angles" between +30º to -30º and an oscillation of 5º for each β – angle was used. An exposure time of 5 s and an average of 3 exposure profiles were used. During *in situ* experiments, the load was continuously recorded. The sample was uniaxially compressed to the target load and stopped. Thereafter, the strain was measured using the *modified Chi mode* which is outlined in detail in supplementary information. The DEC $\left(\frac{1}{2} s_2^{\{hkl\}}\right)$ is thereafter obtained by least square regression fitting of $\left(\frac{\partial \varepsilon_{\beta \chi m}^{\{hkl\}}}{\partial (sin^2 \beta)}\right)$ vs. $\sigma_{11}$ data.

In the Proto iXRD diffractometer, only 2θ angles between 125º-160° can be accessed. Therefore, the list of 2θ angles that are accessible for *in situ* experiments using a particular radiation source is provided in Table 1.

Table 1          2θ angles and corresponding {hkl} planes used for in situ experiments. Also shown is the corresponding radiation source used

| Sample | {hkl} plane | 2θ angle |
|---|---|---|
| **[(MgNiCoCuZn)O]** (Cr $K_\alpha$ radiation) | {113} | 127.05° |
| | {222} | 139.35° |

2.4 Minimization of cost function for estimation of Single Crystal Elastic constants

The outline of the methodology to extract the SECs is shown in Figure 2. To estimate SECs from the experimental data a custom MATLAB code was written.

The experimental data for isotropic elastic constants and DECs are obtained using the methodology outlined in sections 2.2 and 2.3, respectively. It is noted from the literature that the SECs for cubic crystal structure vary between $C_{11} = 4 - 800$ GPa; $C_{12} = 2 - 300$ GPa & $C_{44} = 2 - 300$ GPa [26]. For the various combinations of SECs, the DECs and elastic constants are calculated using the Voigt-Reuss-Hill (VRH) model. More about the Voigt-Reuss-Hill (VRH) model is outlined in the supplementary information.

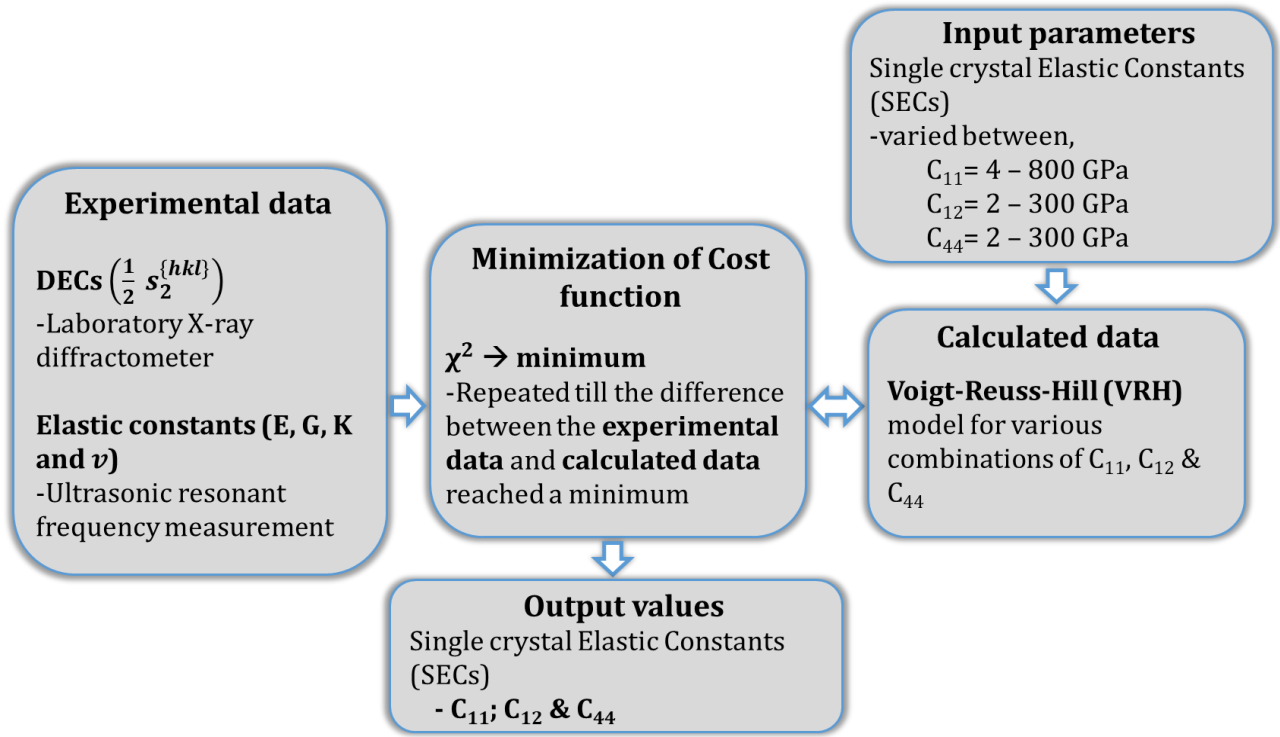

Figure 2     shows the procedure to extract SECs from experimental data by minimization of a cost function

The SECs are subsequently determined by minimizing a cost function between the experimental and calculated values. In the calculated values, the SECs are used as free parameters and are varied over a range. For the [(MgNiCoCuZn)O] sample, the following cost function was used,

$$\chi^2 = \begin{bmatrix} \left\{ \dfrac{\left(\frac{1}{2}\,s_2^{\{113\}}\right)_{\exp} - \left(\frac{1}{2}\,s_2^{\{113\}}\right)_{calc}}{\Delta\left(\frac{1}{2}\,s_2^{\{113\}}\right)_{\exp}} \right\}^2 + \left\{ \dfrac{\left(\frac{1}{2}\,s_2^{\{222\}}\right)_{\exp} - \left(\frac{1}{2}\,s_2^{\{222\}}\right)_{calc}}{\Delta\left(\frac{1}{2}\,s_2^{\{222\}}\right)_{\exp}} \right\}^2 + \\[3mm] \left\{ \dfrac{(E)_{\exp} - (E)_{calc}}{\Delta(E)_{\exp}} \right\}^2 + \left\{ \dfrac{(K)_{\exp} - (K)_{calc}}{\Delta(K)_{\exp}} \right\}^2 + \left\{ \dfrac{(G)_{\exp} - (G)_{calc}}{\Delta(G)_{\exp}} \right\}^2 + \\[3mm] \left\{ \dfrac{(v)_{\exp} - (v)_{calc}}{\Delta(v)_{\exp}} \right\}^2 \end{bmatrix} \quad (1)$$

Where $\Delta$ is the standard deviation in the obtained experimental data.

## 2.5 First principle calculations

In order to validate the experimentally obtained SECs of [(MgNiCoCuZn)O], first principle calculations were performed. The material was modelled with 40 atoms using the Special Quasirandom Structure (SQS) method present in the 'mcsqs' code in the Alloy Theoretic Automated Toolkit (ATAT) [31]. The five cations were randomly arranged at the 4a sites, and oxygen occupied the 4b positions of the rock salt structure. The first principle calculations were performed using the plane wave density functional theory (DFT) as implemented in the Vienna Ab initio Simulation Package (VASP) [32]. The exchange-correlation energy of the Kohn-Sham formulation was approximated using the generalized gradient approximation (GGA) [32] parameterized with Perdew-Bruke-Ernzerhof for solid (PBEsol) [33]. The project augmented wave (PAW) [34] based pseudopotentials were utilized and the following valence electron configurations of the pseudopotentials i.e., Co ($3p^63d^74s^2$), Cu ($3p^63d^{10}4s^1$), Mg ($2p^63s^2$), Ni ($3p^63d^84s^2$), Zn ($3p^63d^{10}4s^2$), and O ($2s^22p^4$) were considered in this work. The plane waves with a cut-off energy of 600 eV and a gamma-centered *k*-point mesh of 8x8x8 were chosen for the convergence of the ground state energy and for the determination of elastic constants. The force criterion was set as 0.01 eV/Å for ionic relaxation and the electronic criterion was specified as $10^{-7}$ eV for the electronic self-consistent loop. The calculations were spin-polarized in nature and the conjugate gradient algorithm [35] was used as the convergence method. To predict the localized nature of the partially filled d-orbitals accurately, the onsite coulombic interactions were considered for Co, and Ni using the DFT+U Hubbard formalism [36]. These U parameters were calculated using the linear response method [37] which was 3.45 and 6.75 for Co and Ni, respectively. The antiferromagnetic nature of rock salt [(MgNiCoCuZn)O] at its ground state was considered by aligning the spin magnetic moments parallel and anti-parallel alternatively, along the (111) plane. In this work, an energy-strain approach method implemented in the Elastic_VASP code [38] was used to determine the elastic constants.

## 3. Results

While the manuscript is restricted to the estimation of single crystal elastic constants (SECs) of the entropy stabilized oxide, where the composition [(MgNiCoCuZn)O] is taken as a prototypical example, but all the initial experiments were performed on polycrystalline nickel to offer robust validation of the proposed methodology. The complete characterization and experimental determination of SECs for nickel can be found in the supplementary information, section 3.

The initial microstructural characterization and phase analysis for the ESO were carried out, followed by which SECs were determined. The single-phase rocksalt type cubic structure for the ESO sample is evident from the X-ray diffractogram shown in Figure 3 (a). The lattice parameter was calculated to be 4.231 Å and matches well with the literature-reported values [3]. The density of the sample was measured using Archimedes' principle and is found to be greater than 99%. Also, as evident from the microstructure in Figure 3 (b), the sample is devoid of any porosity. The average grain size was measured to be $27\pm8$ μm. EDS mapping shown in Figure 3 (c) reveals that the elements are evenly distributed and that there is no evidence of segregation. The sample has a random texture evident from the inverse pole figure shown in Figure 3 (d).

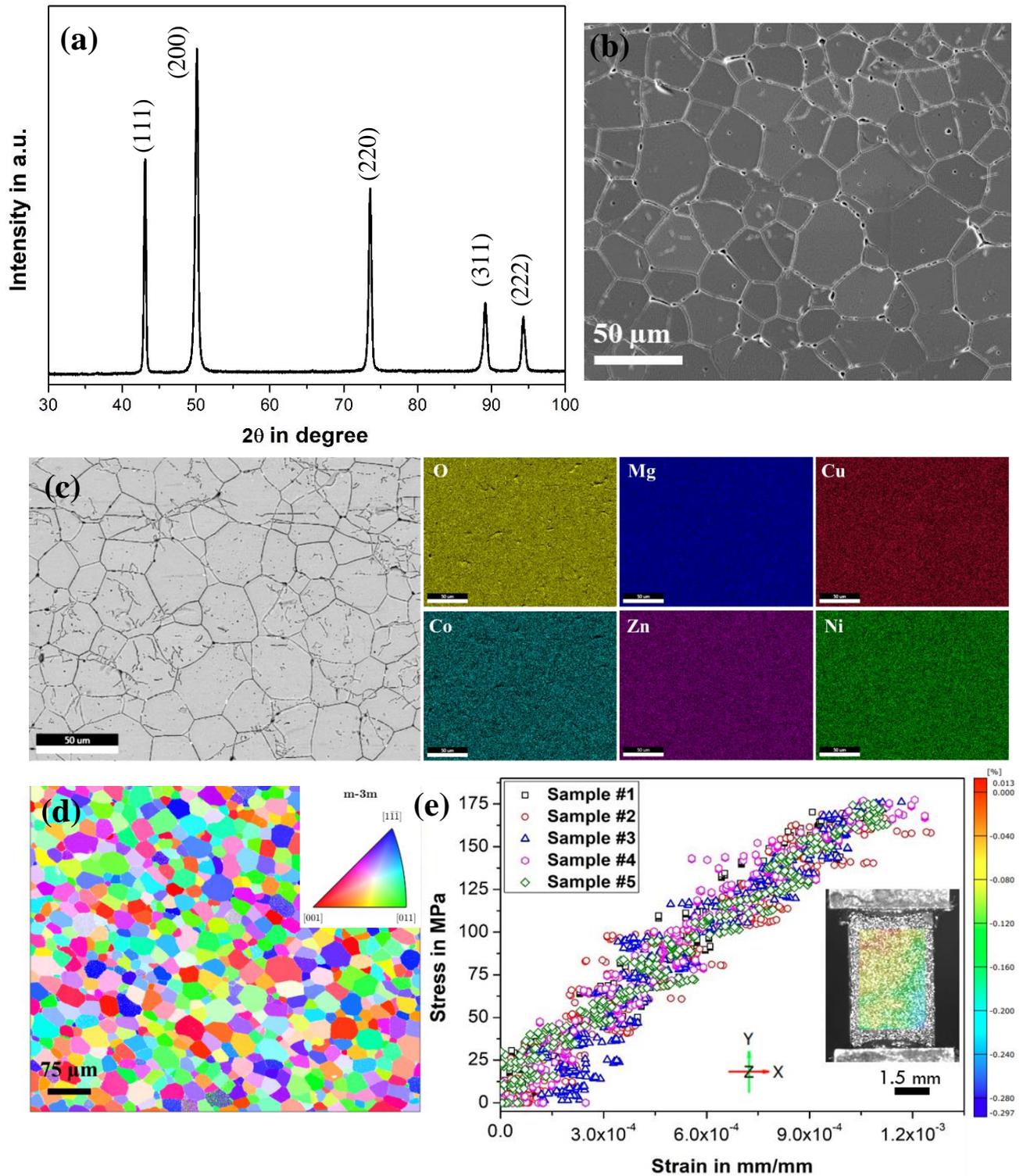

Figure 3    (a) shows the X-ray diffractogram of single phase ESO (b) microstructure of sintered ESO showing little to no porosity, (c) EDS map showing no segregation of any constituent element and (d) inverse pole figure (e) shows the stress-strain data for the ESO sample. Also shown in the inset is the full field strain for compression sample #1 loaded between the compression platens

Table 2        Average EDS compositional analysis showing close to equi-molar atomic percentage for all the constituent metal ions in [(MgNiCoCuZn)O]

| **Element** | Oxygen | Magnesium | Nickel | Cobalt | Copper | Zinc |
|---|---|---|---|---|---|---|
| **Atomic%** | 42.8 ± 3.5 | 10.9 ± 1.4 | 10.0 ± 0.8 | 11.7 ± 0.7 | 12.9 ± 0.9 | 11.7 ± 0.8 |

Table 2 shows the atomic percentage of the constituent elements, and it is observed that the transition metal ions are equiatomic in composition.

Unixial compression was carried out at a strain rate of $4 \times 10^{-4}$ $s^{-1}$. The samples were loaded to a maximum load of 1500 N, beyond which it fractured. Instead of using a strain gauge, DIC was used to measure strain since the strain gauges are not compatible with ceramic samples. Figure 3 (e) shows the stress-strain data obtained from five different samples. Also shown in Figure 3 (e) is the full field strain along the loading direction of the sample. The elastic modulus calculated using uniaxial compression was $140.9 \pm 2.9$ GPa and using ultrasonic measurement was $138.2 \pm 1.1$ GPa. The difference in elastic modulus value is less than 4.5% between the methods. The isotropic elastic constants were determined using ultrasonic resonant frequency testing and two diffraction elastic constants (DECs) using *in situ* loading of samples in a laboratory X-ray diffractometer. The isotropic elastic constants obtained through ultrasonic resonant frequency testing are as follows: Young's modulus (E): $138.2 \pm 0.7$ GPa, shear modulus (G): $51.3 \pm 1.1$ GPa GPa, bulk modulus (K): $150.5 \pm 0.8$ GPa, and Poisson's ratio ($\nu$): $0.347 \pm 0.011$. For DEC measurements, {113} family of planes at 127.05° and {222} family of planes at 139.35° were used. Uniaxial compression coupons of [(MgNiCoCuZn)O] were loaded at equal load intervals of 300 N up to a maximum load of 1500 N, as shown in Figure 4 (a). The lattice strain measurement corresponding to each load interval for the {113} family of planes and {222} family of planes is shown in Figures 4 (b) & 4 (d), respectively. The goodness of fit is above

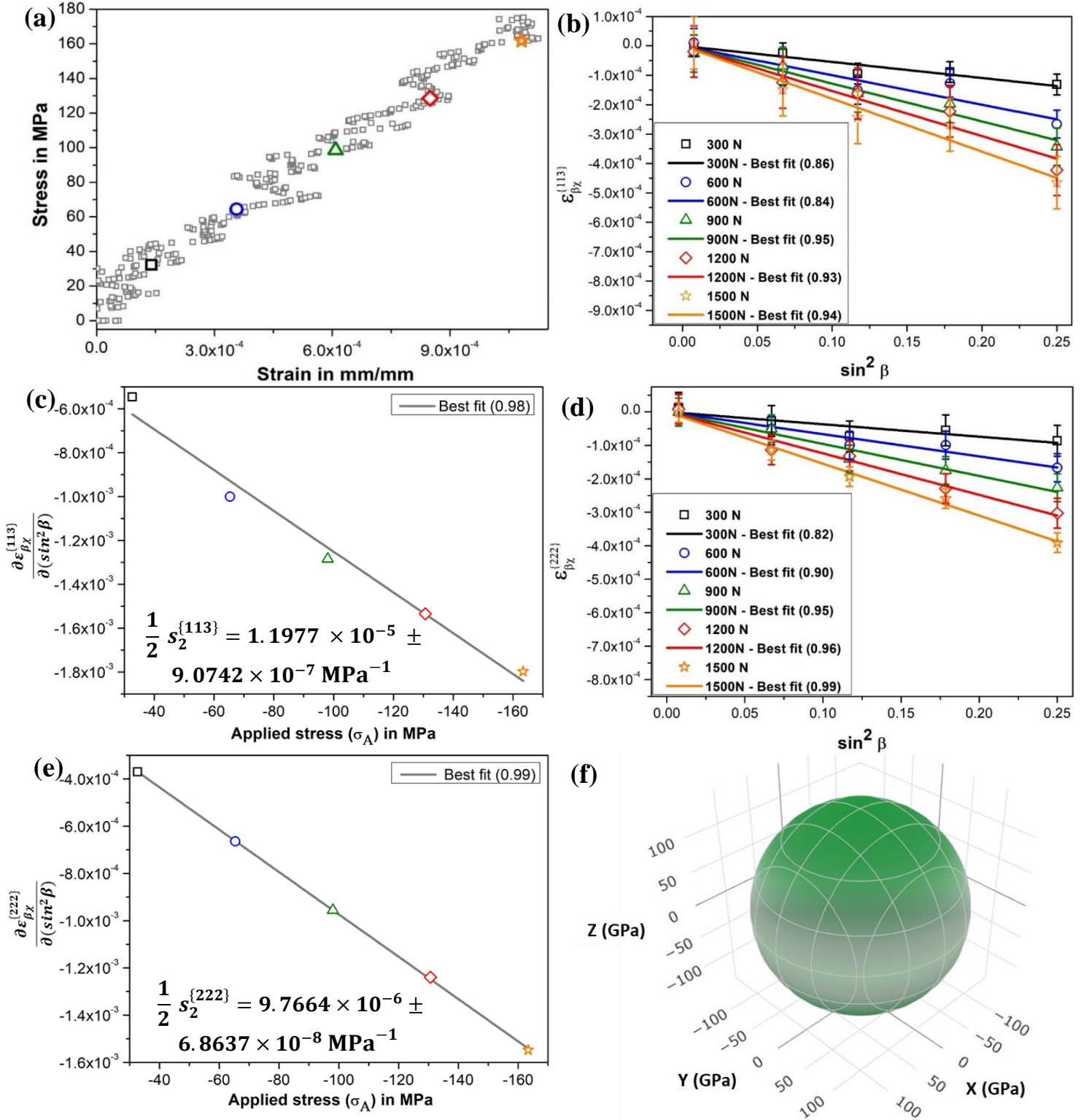

Figure 4    (a) shows the intermittent stress levels (marked with different colors) at which loading was stopped and lattice strain was measured; (b) and (d) shows the lattice strain vs. $sin^2\beta$ data acquired for the {113} & {222} family of planes, respectively at different load interval; (c) and (e) shows the DEC $\left(\frac{1}{2} s_2^{\{113\}}\right)$ & $\left(\frac{1}{2} s_2^{\{222\}}\right)$ calculated from the slope of linear plot of $\frac{\partial \varepsilon_{\beta\chi}^{\{113\}}}{\partial(sin^2\beta)}$ vs. applied stress ($\sigma_A$) & $\frac{\partial \varepsilon_{\beta\chi}^{\{222\}}}{\partial(sin^2\beta)}$ vs. applied stress ($\sigma_A$), respectively; (f) shows the 3D visualization of Young's modulus from calculated SECs 0.84 for all data. The DEC $\left(\frac{1}{2} s_2^{\{hkl\}}\right)$ was calculated to be $1.1977 \times 10^{-5} \pm 9.0742 \times 10^{-7}$ MPa$^{-1}$ for $\{113\}$ family of planes and $9.7664 \times 10^{-6} \pm 6.8637 \times 10^{-8}$ MPa$^{-1}$ for $\{222\}$ family of planes as shown in Figures 4 (c) & 4 (e), respectively. By running a minimization subroutine using the six experimental data (E, K, G, $\boldsymbol{\nu}$, $\frac{1}{2} s_2^{\{113\}}$ & $\frac{1}{2} s_2^{\{222\}}$) and their corresponding relations (refer to equation (1)), SECs were obtained. 3D visualization of Young's modulus calculated from experimentally determined SECs is shown in Figure 4 (f).

### 4. Discussion

The SECs obtained for the sample are given in the table in Figure 5 and are compared with the literature-reported values [15]. To the best of our knowledge, there exist only one work of literature that report the SECs of ESO of composition [(MgNiCoCuZn)O] at ambient conditions. The report is based on a first-principles calculations, where the SECs for various magnetic states were computed [15]. However, considering that this material demonstrates long-range antiferromagnetic behaviour even at room temperature [39,40], only the SECs

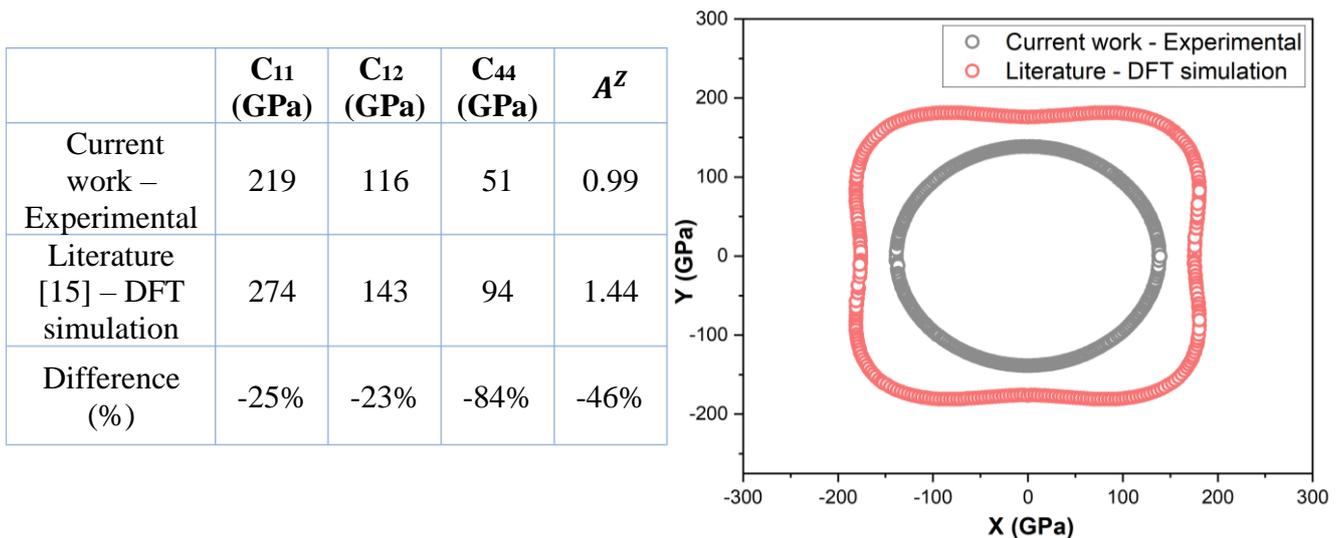

| | $C_{11}$ (GPa) | $C_{12}$ (GPa) | $C_{44}$ (GPa) | $A^Z$ |
|---|---|---|---|---|
| Current work – Experimental | 219 | 116 | 51 | 0.99 |
| Literature [15] – DFT simulation | 274 | 143 | 94 | 1.44 |
| Difference (%) | -25% | -23% | -84% | -46% |

Figure 5    the table shows the measured single crystal elastic constants of ESO compared with the literature-reported values [15]. Also shown is the spatial dependence of Young's modulus along the XY plane calculated from the SECs from current work and literature-reported values calculated for the antiferromagnetic state were used for comparison. The table in Figure 5 also displays the percentage difference for the values of $C_{11}$, $C_{12}$, and $C_{44}$ between the current study and the literature-reported values. Typically, according to the literature, the maximum percentage difference in elastic constants between diffraction techniques and those determined from alternative methods falls within the range of 10 to 13% [21,22,27,41,42]. In this work as well, in the case of nickel, which served as a validation of the proposed methodology, the maximum deviation observed in elastic constants was within 11.5%, as illustrated in the table within Figure S4. However, in the case of [(MgNiCoCuZn)O], the observed deviation is considerably higher. Consequently, a comprehensive discussion of the estimated SECs is essential to discern and comprehend the underlying differences.

The spatial variation of Young's modulus across the XY plane was computed using the SECs from the present study and values reported in the literature. The results are depicted in Figure 5. A noticeable observation is that the spatial distribution of Young's modulus appears isotropic in the current study, whereas variations are evident in Young's modulus calculated from SECs reported in the literature. To ascertain this further Zenner anisotropy ratio ($A^Z$) was calculated using the formulation, $A^Z = 2C_{44}/(C_{11} - C_{12})$, and the results are presented in the table within Figure 5 for both the SECs obtained from the current study and those reported in the literature. It is evident that for the SECs derived from the present work, $A^Z$ is in close proximity to 1, indicating a high degree of isotropy in the material. In contrast, for the literature-reported SECs, $A^Z$ is measured at 1.44. From the literature, there is only a single existing experimental study on measuring SECs for these ESO [16]. Even in this particular study, SECs were determined starting from pressures of 1.2 GPa, and the resulting $A^Z$ was calculated to be around 4.9. This value is considerably higher than the $A^Z$ values derived from the results of this study. Therefore, to explore this aspect further, indentation mapping was conducted to assess whether the variation in indentation modulus with respect to crystallographic orientation remains constant or changes. Figure 6 (a) shows the inverse pole figure after indentation mapping. The variation of the indentation modulus as a function of crystallography orientation is shown in Figure 6 (b). In this plot indentations within individual grains are considered, while all indentations occurring along grain boundaries are excluded. Given the limitation of grain size only one indentation per grain could be conducted and hence Figure 6 (b) does not include error bars for the reported indentation modulus. Also, shown in Figure 6 (b) is the indentation modulus as a function of the orientation factor calculated from SECs from current work and literature reported values. The indentation modulus from SECs is calculated using the relation from Vlassak and Nix [43]. The crystallographic orientation {$hkl$} dependent elastic modulus is given by

$$M_{hkl} = 1.058\beta_{hkl} \left(\frac{E}{1-\nu^2}\right)_{isotropic} \tag{2}$$

where, $\beta_{hkl}$ is the orientation factor and for crystals of cubic symmetry it is given by,

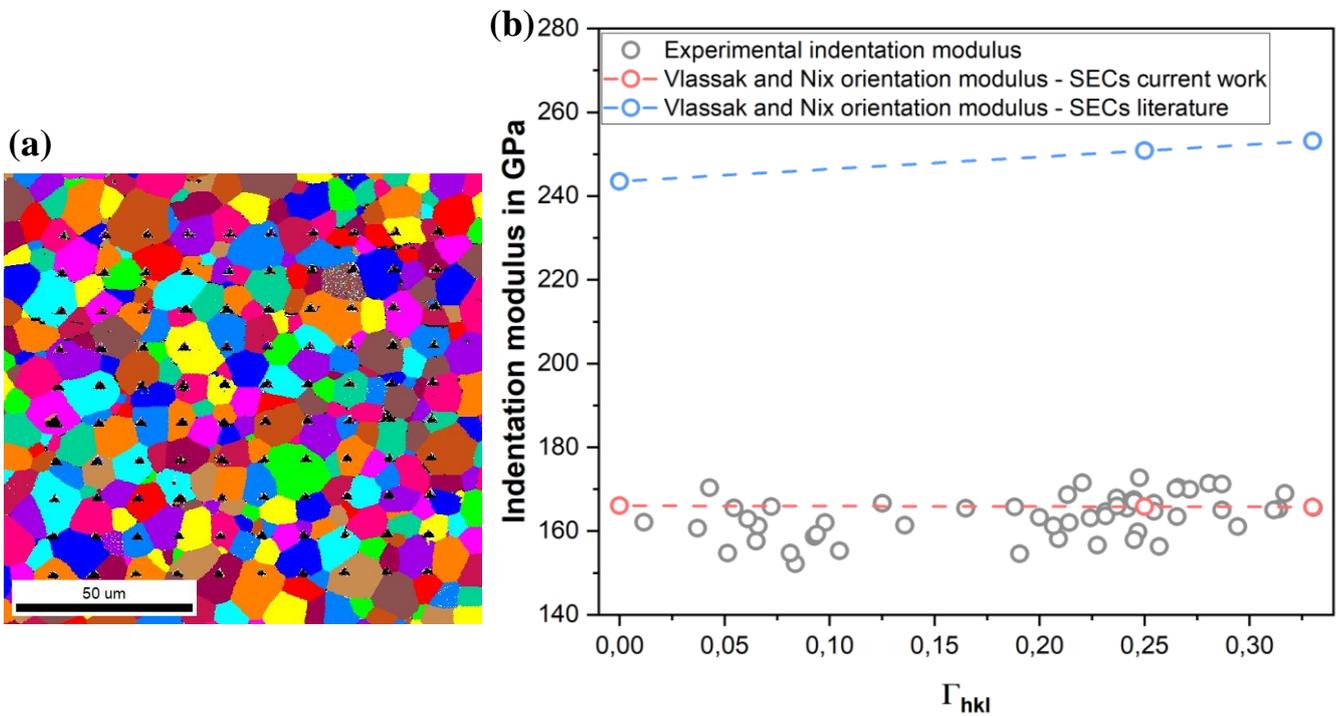

Figure 6      (a) shows the inverse pole figure of the ESO sample post-indentation mapping, (b) shows the variation in indentation modulus with the orientation factor. Additionally, it shows the orientation-dependent modulus calculated using the Vlassak and Nix relation for the SECs obtained in this experimental study and from the literature

$$\beta_{hkl} = a + c(A^Z - A_0)^B \qquad (3)$$

where $a$, $c$, $A_0$ and $B$ depend on Poisson's ratio along the cube's directions and are given in [43]. The analysis of the experimental data indicates a consistent indentation modulus across various orientations. These values align closely with the orientation modulus calculated using SECs derived in this study. Conversely, when utilizing SECs from existing literature, the calculated orientation modulus demonstrates significant discrepancies both in absolute value and in trend.

In cubic crystals with centrosymmetry, the elastic tensor typically displays anisotropic properties. However, exceptions exist, as seen in tungsten, where its isotropic behavior is attributed to electronic structure [44]. Similarly, despite the face-centered cubic structure with centrosymmetry (Fm$\bar{3}$m), [(MgNiCoCuZn)O] exhibits high isotropy in experimental results. Conversely, literature-based density functional theory (DFT) simulations reveal anisotropy in this material, with a Zener ratio of 1.44 [15]. In the same study, various magnetic states of [(MgNiCoCuZn)O] were analyzed for Zener ratios, resulting in values of 1.12 and 0.98 for paramagnetic and non-magnetic states, respectively. While the non-magnetic state displayed notable isotropy, it was noted that local distortions arising from Cu and Zn ions could be contributing factors. These local distortions were ascertained using a radial distribution function. For further validation DFT simulations were carried out in this work. In the calculations, only the antiferromagnetic state of [(MgNiCoCuZn)O] was considered. To enhance computational speed, the study utilized a supercell consisting of 40 atoms and employed the energy-strain method for calculating elastic constants. This approach differed from the previously reported literature, where 160 atoms and the stress-strain method were utilized for the same purpose [15]. Also, the U parameter in the present DFT calculations were obtained using a linear response method when compared to an arbitrary value chosen in the work reported in literature. The table in Figure 7 shows the comparison of elastic constants

| | $C_{11}$ (GPa) | $C_{12}$ (GPa) | $C_{44}$ (GPa) | $A^Z$ |
|---|---|---|---|---|
| Current work – DFT simulation | 311 | 151 | 84 | 1.05 |
| Literature [15] – DFT simulation | 274 | 143 | 94 | 1.44 |
| Difference (%) | 12% | 5% | -12% | -37% |

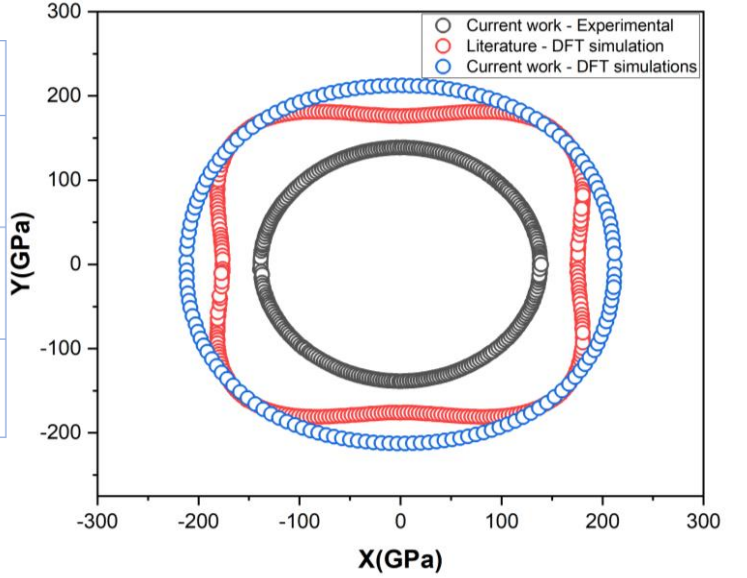

Figure 7    the table shows the DFT calculated single crystal elastic constants of ESO compared with the literature-reported values [15]. Also shown is the spatial dependence of Young's modulus along the XY plane calculated from the SECs from the current work (both experimental and DFT calculations) and literature-reported values between the current and literature reported DFT simulations. A maximum of $\pm 12\%$ deviation is observed between the values and also shown in Figure 7 is the spatial dependence of Young's modulus along the XY plane which shows that the values from the current DFT simulations are in good agreement with the literature values. A notable distinction is that the Zener ratio obtained from our DFT simulations is close to 1, in contrast to the results available in literature. To confirm that the isotropic behaviour is not a consequence of local lattice distortions, contrary to what was documented for the non-magnetic state of [(MgNiCoCuZn)O] in the literature, we calculated the radial distribution function for the converged [(MgNiCoCuZn)O] structure. As depicted in Figure 8 (a), the radial distribution of the converged structure closely aligns with the ideal rocksalt structure. Therefore, we believe, the primary factor contributing to the variance in the Zener ratio between our simulations and the literature could be the U parameter utilized in the Hubbard formalism for modelling transition metal systems. It is established that the U parameter substantial influences electronic states and consequently impacts the calculated properties in simulations, particularly for transition metal oxides [45,46]. Considering that $Cu^{2+}$ ions in [(MgNiCoCuZn)O] already exhibit Jahn-Teller distortion, which is known to influence the electronic state and thereby the material properties [13,47], obtaining an accurate estimation of the U parameter becomes crucial for predicting

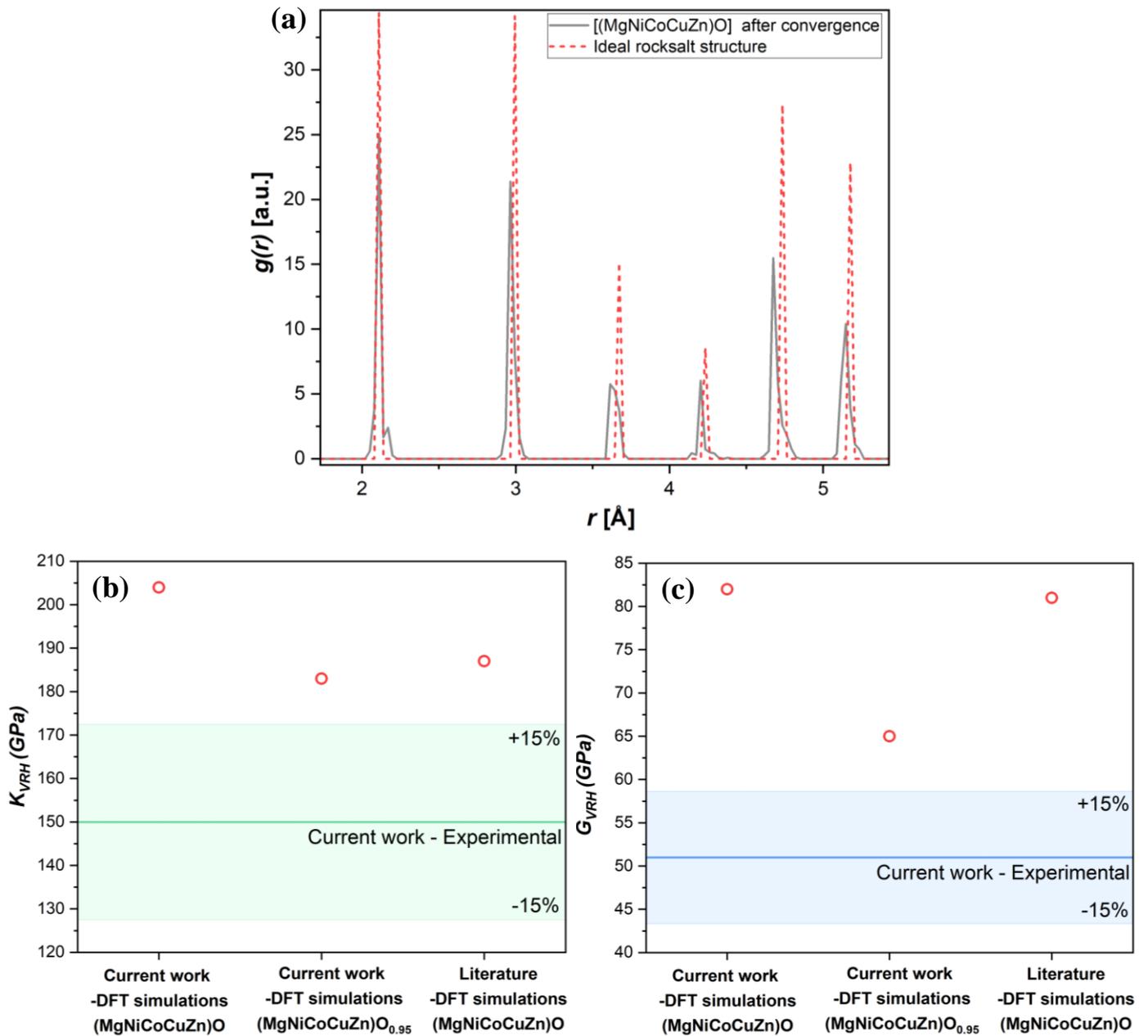

Figure 8      (a) shows the radial distribution of the converged [(MgNiCoCuZn)O] structure along with the ideal rocksalt structure and (b) and (c) shows the bulk modulus ($K_{VRH}$) and shear modulus ($G_{VRH}$) respectively, calculated using the Voigt-Reuss-Hill average for the SECs obtained using DFT calculations. Also, shown in graph (b) and (c) is the reference $K_{VRH}$ and $G_{VRH}$ along with 15% standard deviation obtained using SECs from current experiments.

elastic constants in DFT simulations. Even after accounting for the discrepancies in the anisotropy ratio the differences in absolute values still require an explanation. In existing literature, it has been noted that across various compounds, a maximum deviation of approximately $\pm 15\%$ exists between the bulk and shear moduli calculated via DFT and those obtained experimentally through the Voigt-Reuss-Hill average utilizing single crystal elastic constants [48]. However, in this case a deviation of more than 25% was observed. The Young's modulus, bulk modulus, and shear modulus, calculated from the Voigt-Reuss-Hill averaging method using single crystal elastic constants obtained from both the present study and previous literature, are detailed in Table S1, including corresponding standard deviations.

One plausible hypothesis accounting for the significant disparities between experimental and Density Functional Theory (DFT) computed Single Crystal Elastic Constants (SECs) is the potential non-stoichiometry of the [(MgNiCoCuZn)O] ESO composition. Specifically, at room temperature, it is documented that this composition hosts oxygen vacancies within the lattice [14]. Researchers have indeed confirmed the presence of oxygen vacancies up to a concentration of 0.07 under low oxygen partial pressures [49]. Given that presence of oxygen vacancies could affect the electronic structure, DFT calculations were conducted for an ESO composition $(MgNiCoCuZn)O_{0.95}$, wherein the oxygen lattice was simulated to have a vacancy fraction of 0.05. The DFT calculated values for this composition are given Table S1. One interesting aspect is that the absolute values of SECs reduce when compared to stoichiometric $(MgNiCoCuZn)O$ and also the values approach close to the experimental results. The shear modulus and bulk modulus were also calculated using the Voigt-Reuss-Hill average and the graphical summary of results compared with the experimental results are shown in Figure 8 (b) and (c) respectively. From DFT calculations it can be clearly observed that the precise estimation of Single Crystal Elastic Constants (SECs) relies significantly on the electronic structure of the system. It is therefore essential that a reasonable estimate for all the parameters in DFT calculations be used.

5. Conclusions:

The following conclusions can be drawn from the present work.

- The single crystal elastic constants (SECs) were determined using a robust micromechanical model (Voigt-Reuss-Hill model) along with experimental data and an optimization subroutine. Proof of concept experiments were carried out on commercial purity polycrystalline nickel samples. In total, six experimental data were used to estimate SECs, 2 DECs and four isotropic elastic constants (Young's modulus (E), shear modulus (G), bulk modulus (K) and Poisson's ratio ($v$)). The DECs were obtained by *in situ* uniaxial loading of polycrystalline samples. The four elastic constants were measured using ultrasonic resonant frequency testing of the bulk samples. From the minimization of a cost function, the SECs for the nickel samples were estimated and there was good agreement between the experimentally determined and the literature reported SECs value.

- With the methodology to determine SECs validated for nickel, experiments were carried out for the entropy stabilized oxides of composition [(MgNiCoCuZn)O]. All the experimental conditions were kept identical except that the DECs for ESO was measured from the {113} and {222} family of planes. The SECs were estimated to be 219 GPa, 116 GPa and 51 GPa for $C_{11}$, $C_{12}$ and $C_{44,}$ respectively. The deviation in the experimental results with the literature reported values were between 25%, 23% and 84% for $C_{11}$, $C_{12}$ and $C_{44,}$ respectively. The literature reported SECs for this system were estimated using DFT calculations.

- For validation purposes, experimentally derived SECs were utilized to compute the orientation-dependent elastic modulus using the Vlassak and Nix relation. This was then compared with the indentation modulus obtained as a function of crystallographic orientation. The comparison showed strong agreement, instilling confidence in the reliability of the experimentally determined SECs. This necessitated a need to look at DFT simulation and understand the discrepancies in literature values.

- DFT simulations revealed a strong interdependency on U parameter used for modelling the transition metal compounds plays a critical role in the electronic properties of the structures and thereby affecting the absolute values of SEC determined. Experiments and simulations also revealed that entropy stabilized oxides of composition [(MgNiCoCuZn)O] is isotropic despite its crystal structure of face-centered cubic structure with centrosymmetry. The observed isotropy is strongly linked to its electronic properties.

**Acknowledgements**

The authors would like to thank the Center for Non-Destructive Evaluation, IIT Madras for providing access to ultrasonic frequency resonant testing rig. The funding received from the Institute of Eminence Research Initiative Project on Materials and Manufacturing for Futuristic Mobility (project no. SB20210850MMMHRD008275) and the financial support from the Exploratory Research Project, IC&SR, IIT Madras, India (Grant No. MET 17-18 850 RFER RAVK) is gratefully acknowledged.

**References**

[1]  D.B. Miracle, O.N. Senkov, A critical review of high entropy alloys and related concepts, Acta Mater. 122 (2017) 448–511. https://doi.org/10.1016/j.actamat.2016.08.081.

[2]  E.P. George, D. Raabe, R.O. Ritchie, High-entropy alloys, Nat. Rev. Mater. 4 (2019) 515–534. https://doi.org/10.1038/s41578-019-0121-4.

[3]  C.M. Rost, E. Sachet, T. Borman, A. Moballegh, E.C. Dickey, D. Hou, J.L. Jones, S.

Curtarolo, J. Maria, Entropy-stabilized oxides, Nat. Commun. (2015).
https://doi.org/10.1038/ncomms9485.

[4]     J. Zhang, X. Zhang, Y. Li, Q. Du, X. Liu, X. Qi, High-entropy oxides 10La2O3-
        20TiO2-10Nb2O5-20WO3-20ZrO2 amorphous spheres prepared by containerless
        solidificatio, Mater. Lett. 244 (2019) 167–170.
        https://doi.org/10.1016/j.matlet.2019.01.017.

[5]     A. Sarkar, R. Djenadic, D. Wang, C. Hein, R. Kautenburger, Rare earth and transition
        metal based entropy stabilised perovskite type oxides, J. Eur. Ceram. Soc. 38 (2018)
        2318–2327. https://doi.org/10.1016/j.jeurceramsoc.2017.12.058.

[6]     J. Da, M. Stygar, A. Mikuła, A. Knapik, K. Mroczka, W. Tejchman, M. Danielewski,
        M. Martin, Synthesis and microstructure of the ( Co , Cr , Fe , Mn , Ni ) 3 O 4 high
        entropy oxide characterized by spinel structure, Mater. Lett. 216 (2018) 32–36.
        https://doi.org/10.1016/j.matlet.2017.12.148.

[7]     J. Gild, M. Samiee, L. Braun, T. Harrington, H. Vega, High-entropy fluorite oxides, J.
        Eur. Ceram. Soc. 38 (2018) 3578–3584.
        https://doi.org/10.1016/j.jeurceramsoc.2018.04.010.

[8]     A. Giri, J.L. Braun, P.E. Hopkins, Reduced dependence of thermal conductivity on
        temperature and pressure of multi-atom component crystalline solid solutions, J. Appl.
        Phys. 123 (2018). https://doi.org/10.1063/1.5010337.

[9]     J.L. Braun, C.M. Rost, M. Lim, A. Giri, D.H. Olson, G.N. Kotsonis, G. Stan, D.W.
        Brenner, J.P. Maria, P.E. Hopkins, Charge-Induced Disorder Controls the Thermal
        Conductivity of Entropy-Stabilized Oxides, Adv. Mater. 1805004 (2018) 1–8.
        https://doi.org/10.1002/adma.201805004.

[10]  A. Sarkar, Q. Wang, A. Schiele, M.R. Chellali, S.S. Bhattacharya, D. Wang, T. Brezesinski, H. Hahn, L. Velasco, B. Breitung, High-Entropy Oxides : Fundamental Aspects and Electrochemical Properties, Adv. Mater. 31 (2019) 1806236-1–9. https://doi.org/10.1002/adma.201806236.

[11]  C. Oses, High-entropy ceramics, Nat. Rev. Mater. 5 (2020) 295–309. https://doi.org/10.1038/s41578-019-0170-8.

[12]  V. Nallathambi, L. Kumar, D. Wang, A.A. Naberezhnov, S. V Sumnikov, E. Ionescu, R. Kumar, Tuning the mechanical and thermal properties of ( MgNiCoCuZn ) O by intelligent control of cooling rates, J. Eur. Ceram. Soc. 43 (2023) 4517–4529. https://doi.org/10.1016/j.jeurceramsoc.2023.03.016.

[13]  Z. Rák, J.P. Maria, D.W. Brenner, Evidence for Jahn-Teller compression in the (Mg Co Ni Cu Zn)O entropy-stabilized oxide: A DFT study, Mater. Lett. 217 (2018) 300–303. https://doi.org/10.1016/j.matlet.2018.01.111.

[14]  L.K. Bhaskar, V. Nallathambi, R. Kumar, Critical role of cationic local stresses on the stabilization of entropy-stabilized transition metal oxides, J. Am. Ceram. Soc. 103 (2020) 3416–3424. https://doi.org/10.1111/jace.17029.

[15]  A.E. Pitike, Krishna Chaitanya Marquez-rossy, A. Flores-betancourt, D.X. Chen, E. Lara-curzio, V.R. Cooper, On the elastic anisotropy of the entropy- stabilized oxide ( Mg , Co , Ni , Cu , Zn ) O compound, J. Appl. Phys. 015101 (2020). https://doi.org/10.1063/5.0011352.

[16]  B. Yue, W. Dai, X. Zhang, H. Zhang, W. Zhong, Deformation behavior of high-entropy oxide ( Mg , Co , Ni , Cu , Zn ) O under extreme compression, Scr. Mater. 219 (2022) 114879. https://doi.org/10.1016/j.scriptamat.2022.114879.

[17]   H. Seiner, P. Sedlák, Resonant ultrasound spectroscopy, J. Phys. Condens. Matter. 9 (1997) 6001–6029.

[18]   P.E. Aba-Perea, T. Pirling, P.J. Withers, J. Kelleher, S. Kabra, M. Preuss, Determination of the high temperature elastic properties and diffraction elastic constants of Ni-base superalloys, Mater. Des. 89 (2016) 856–863. https://doi.org/10.1016/j.matdes.2015.09.152.

[19]   A. Cervellino, P.M. Derlet, H. Van Swygenhoven, Elastic properties determined from in situ X-ray diffraction, Acta Mater. 54 (2006) 1851–1856. https://doi.org/10.1016/j.actamat.2005.12.022.

[20]   A.J.G. Lunt, M.Y. Xie, N. Baimpas, S.Y. Zhang, S. Kabra, J. Kelleher, T.K. Neo, A.M. Korsunsky, Calculations of single crystal elastic constants for yttria partially stabilised zirconia from powder diffraction data, J. Appl. Phys. 116 (2014). https://doi.org/10.1063/1.4891714.

[21]   E.H. Kisi, C.J. Howard, Elastic constants of tetragonal zirconia measured by a new powder diffraction technique, J. Am. Ceram. Soc. 81 (1998) 1682–1684. https://doi.org/10.1111/j.1151-2916.1998.tb02533.x.

[22]   S. Fréour, D. Gloaguen, M. François, A. Perronnet, R. Guillén, Determination of single-crystal elasticity constants in a cubic phase within a multiphase alloy: X-ray diffraction measurements and inverse-scale transition modelling, J. Appl. Crystallogr. 38 (2005) 30–37. https://doi.org/10.1107/S0021889804023441.

[23]   M.F. Slim, A. Alhussein, E. Zgheib, M. François, Determination of single-crystal elasticity constants of the beta phase in a multiphase tungsten thin film using impulse excitation technique, X-ray diffraction and micro-mechanical modeling, Acta Mater. 175 (2019) 348–360. https://doi.org/10.1016/j.actamat.2019.06.035.

[24]    A. Heldmann, M. Hoelzel, M. Hofmann, W. Petry, E. Griesshaber, Diffraction-based determination of single-crystal elastic constants of polycrystalline titanium alloys, J. Appl. Crystallogr. 52 (2019).

[25]    A.J.G. Lunt, M.Y. Xie, N. Baimpas, S.Y. Zhang, S. Kabra, J. Kelleher, T.K. Neo, A.M. Korsunsky, Calculations of single crystal elastic constants for yttria partially stabilised zirconia from powder diffraction data, J. Appl. Phys. 116 (2014). https://doi.org/10.1063/1.4891714.

[26]    X. Du, J.C. Zhao, Facile measurement of single-crystal elastic constants from polycrystalline samples, Npj Comput. Mater. 3 (2017) 1–7. https://doi.org/10.1038/s41524-017-0019-x.

[27]    J.R. Neighbours, F.W. Bratten, C.S. Smith, The elastic constants of nickel, J. Appl. Phys. 23 (1952) 389–393. https://doi.org/10.1063/1.1702218.

[28]    L. Kumar, J. Rapp, A. Nandi, A. Kumar, K. Zahir, B. Koohbor, Out-of-oven rapid synthesis of entropy stabilized oxides using radio frequency heating, J. Mater. Res. Technol. 24 (2023) 1150–1161. https://doi.org/10.1016/j.jmrt.2023.03.060.

[29]    W.C. Oliver, G.M. Pharr, An improved technique for determining hardness and elastic modulus using load and displacement sensing indentation experiments, J. Mater. Res. 7 (1992).

[30]    L.K. Bhaskar, G. Kumar, N. Srinivasan, R. Kumar, Design and development of a miniaturized multiaxial test setup for in situ x-ray diffraction experiments Design and development of a miniaturized multiaxial test setup for in situ x-ray diffraction experiments, Rev. Sci. Instrum. 015116 (2021). https://doi.org/10.1063/5.0031805.

[31]    A. Van De Walle, P. Tiwary, M. De Jong, D.L. Olmsted, M. Asta, A. Dick, D. Shin,

Y. Wang, L. Chen, Z. Liu, M. Carlo, CALPHAD : Computer Coupling of Phase Diagrams and Thermochemistry Ef fi cient stochastic generation of special quasirandom structures, Calphad. 42 (2013) 13–18. https://doi.org/10.1016/j.calphad.2013.06.006.

[32]  R. Hafner, Ab-Initio Simulations of Materials Using VASP : Density-Functional Theory and Beyond, J. Comput. Chem. 29 (2008). https://doi.org/10.1002/jcc.

[33]  I. Csonka, O.A. Vydrov, G.E. Scuseria, L.A. Constantin, J.P. Perdew, A. Ruzsinszky, X. Zhou, K. Burke, Restoring the Density-Gradient Expansion for Exchange in Solids and Surfaces, Phys. Rev. Lett. 100 (2008) 1–4. https://doi.org/10.1103/PhysRevLett.100.136406.

[34]  G. Kresse, D. Joubert, From ultrasoft pseudopotentials to the projector augmented-wave method, Phys. Rev. B. 59 (1999) 11–19.

[35]  W.T. Vetterling, S.A. Teukolsky, B.P. Flannery, W.H. Press, Numerical Recipes in C The Art of Scientific Computing, Cambridge: Cambridge university press, 1992.

[36]  S.L. Dudarev, G.A. Botton, S.Y. Savrasov, C.J. Humphreys, A.P. Sutton, Electron-energy-loss spectra and the structural stability of nickel oxide: An LSDA + U study, Phys. Rev. B. 57 (1998) 1505–1509.

[37]  M. Cococcioni, S. De Gironcoli, Linear response approach to the calculation of the effective interaction parameters in the LDA+ U method, Phys. Rev. B. 71 (2005) 1–16. https://doi.org/10.1103/PhysRevB.71.035105.

[38]  P. Kumar, I. Adlakha, Effect of Interstitial Hydrogen on Elastic Behavior of Metals : An Ab-Initio Study, J. Eng. Mater. Technol. 145 (2023). https://doi.org/10.1115/1.4055097.

[39]  J. Zhang, J. Yan, S. Calder, Q. Zheng, M.A. Mcguire, D.L. Abernathy, Y. Ren, S.H. Lapidus, K. Page, H. Zheng, J.W. Freeland, J.D. Budai, R.P. Hermann, Long-Range Antiferromagnetic Order in a Rocksalt High Entropy Oxide, Chem. Mater. 31 (2019) 3705−3711. https://doi.org/10.1021/acs.chemmater.9b00624.

[40]  A. Phys, T. Takayama, D. Bérardan, A. Hoser, H. Takagi, N. Dragoe, Long-range magnetic ordering in rocksalt- type high-entropy oxides Long-range magnetic ordering in rocksalt-type high-entropy oxides, Appl. Phys. Lett. 114 (2019) 1–6. https://doi.org/10.1063/1.5091787.

[41]  T. GnaÈupel-Herold, P.C. Brand, H.J. Prask, Calculation of Single-Crystal Elastic Constants for Cubic Crystal Symmetry from Powder Diffraction Data, J. Appl. Crystallogr. 31 (1998) 929–935. https://doi.org/10.1107/S002188989800898X.

[42]  A. Heldmann, M. Hoelzel, M. Hofmann, W. Gan, W.W. Schmahl, E. Griesshaber, T. Hansen, N. Schell, W. Petrya, Diffraction-based determination of single-crystal elastic constants of polycrystalline titanium alloys, J. Appl. Crystallogr. 52 (2019) 1144–1156. https://doi.org/10.1107/S1600576719010720.

[43]  J.J. Vlassak, N.W. D, Measuring the elastic properties of anisotropic materials by means of indenatation experiments, J. Mech. Phys. Solids. 42 (1994) 1223–1245.

[44]  M. Lazar, E. Agiasofitou, T. Böhlke, Mathematical modeling of the elastic properties of cubic crystals at small scales based on the Toupin – Mindlin anisotropic first strain gradient elasticity, Contin. Mech. Thermodyn. 34 (2022) 107–136. https://doi.org/10.1007/s00161-021-01050-y.

[45]  K. Yu, E.A. Carter, Communication: Comparing ab initio methods of obtaining effective U parameters for closed-shell materials, J. Chem. Phys. 121105 (2014). https://doi.org/10.1063/1.4869718.

[46] D.S. Lambert, D.D. O'Regan, Use of DFT + U + J with linear response parameters to predict non-magnetic oxide band gaps with hybrid-functional accuracy, Phys. Rev. Res. 013160 (2023). https://doi.org/10.1103/PhysRevResearch.5.013160.

[47] J. Yan, L. Zhang, J. Liu, N. Li, N. Tamura, B. Chen, W.L. Mao, Pressure-induced suppression of Jahn – Teller distortions and enhanced electronic properties in high-entropy oxide Pressure-induced suppression of Jahn – Teller distortions and enhanced electronic properties in, Appl. Phys. Lett. 151901 (2021) 0–6. https://doi.org/10.1063/5.0067432.

[48] M. De Jong, W. Chen, T. Angsten, A. Jain, R. Notestine, A. Gamst, M. Sluiter, C.K. Ande, S. Van Der Zwaag, J.J. Plata, C. Toher, S. Curtarolo, G. Ceder, K.A. Persson, M. Asta, Charting the complete elastic properties of inorganic crystalline compounds, Sci. Data. 2 (2015) 1–13. https://doi.org/10.1038/sdata.2015.9.

[49] Z. Grzesik, G. Smoła, M. Stygar, J. Dąbrowa, M. Zajusz, K. Mroczka, M. Danielewski, Defect structure and transport properties in (Co,Cu,Mg,Ni,Zn)O high entropy oxide, J. Eur. Ceram. Soc. 39 (2019) 4292–4298. https://doi.org/10.1016/j.jeurceramsoc.2019.06.018.